\documentstyle[12pt]{article}

\begin{document}

\topmargin -30pt
\addtolength{\textheight}{130pt}
\setlength{\textwidth}{16cm}
\setlength{\topmargin}{-1.3cm}
\setlength{\evensidemargin}{0.5cm}
\setlength{\oddsidemargin}{0.5cm}

\title{The ac-Driven Motion of Dislocations in a Weakly Damped Frenkel-Kontorova
Lattice}
\author{Giovanni Filatrella$^1$ and Boris A. Malomed$^2$}
\date{$^1$Unit\`{a} INFM Salerno and Facolt\`a di Scienze, Universit\`a del
Sannio, Via Caio Ponzio Telesino 11, I-82100 Benevento, Italy; \\
e-mail: giofil@physics.unisa.it \\
$^2$Department of Interdisciplinary Studies, Faculty of Engineering, Tel
Aviv University, Tel Aviv 69978, Israel; \\
e-mail malomed@eng.tau.ac.il \\ \vspace{1cm}
June 22, 1999}
\maketitle

%\newpage
{\bf {ABSTRACT} }

By means of numerical simulations, we demonstrate that ac field can support
stably moving collective nonlinear excitations in the form of dislocations
(topological solitons, or kinks) in the Frenkel-Kontorova (FK) lattice with
weak friction, which was qualitatively predicted by
Bonilla and Malomed [Phys. Rev. B{\bf \ 43}, 11539 (1991)]. Direct
generation of the moving dislocations turns out to be virtually impossible;
however, they can be generated initially in the lattice subject to an
auxiliary spatial modulation of the on-site potential strength. Gradually
relaxing the modulation, we are able to get the stable moving dislocations
in the uniform FK lattice with the periodic boundary conditions, provided
that the driving frequency is close to the gap frequency of the linear
excitations in the uniform lattice. The excitations that can be generated
this way have a large and noninteger index of commensurability with the
lattice (so suggesting that the actual value of the commensurability index
is irrational). The simulations reveal two different types of the moving
dislocations: broad ones, that extend, roughly, to half the full length of
the periodic lattice (in that sense, they cannot be called solitons), and
localized soliton-like dislocations, that can be found in an excited state,
demonstrating strong persistent internal vibrations. The minimum (threshold)
amplitude of the driving force necessary to support the traveling excitation
is found as a function of the friction coefficient. Its extrapolation
suggests that the threshold does not vanish at the zero friction, which may
be explained by radiation losses. The moving dislocation can be observed
experimentally in an array of coupled small Josephson junctions in the form
of an {\it inverse Josephson effect}, i.e., a dc-voltage response to the
uniformly applied ac bias current.

%\newpage

\section{INTRODUCTION}

The role of solitons (localized nonlinear collective excitations) in
nonlinear dynamical models of solid-state and polymer physics is well known
(see, e.g., \cite{general,review}). In real systems, the most essential
problem is compensation of dissipative losses. In particular, the losses
always give rise to a friction force acting on a moving soliton. The
friction is usually balanced by a driving force, which may be induced by an
external field. If the losses are weak enough, i.e., the corresponding model
is {\em underdamped} (which is typical for the Josephson junctions \cite{MS}
), the necessary driving field is also weak. In uniform continual systems
with friction, the external drive cannot support progressive (on average)
soliton's motion unless the field contains a dc component. However, an ac
drive with {\em zero} dc component can give rise, in the presence of
friction, to motion of a soliton at a nonzero mean velocity in a continual
system periodically modulated in space \cite{Bob}, or in a discrete lattice.
A physically relevant modulated continual model, in which this effect was
studied in detail analytically and numerically, describes a long Josephson
junction with a periodically modulated critical current density (which can
be easily realized in the experiment by means of periodic modulation of the
thickness of the junction's dielectric layer \cite{Ustinov}):
\begin{equation}
\phi _{tt}-\phi _{xx}+\left[ 1+\epsilon \sin (x/L)\right] \sin \phi =
-\alpha \phi _t+\gamma \sin (\omega t)\,,  \label{JJ}
\end{equation}
where $\phi $ is the magnetic flux inside the junction, the coordinate $x$
and time $t$ are measured in units of the Josephson penetration length and
inverse gap frequency of the junction, $\alpha $ is the dissipative
constant, $\epsilon $ and $L$ are the amplitude and period of the
modulation, $\gamma $ and $\omega $ being the amplitude and frequency of the
ac bias current driving the junction. A soliton in the model (\ref{JJ})
moving at a nonzero mean velocity represents an {\em inverse} Josephson
effect.

The basic mechanism, which is common to the modulated continual model (\ref
{JJ}) and the discrete models described below, is that the motion of a
topological soliton (kink) with a nonzero mean velocity $v$ may be possible,
in the lossy system, in a {\em resonant} case, when the period $2\pi /\omega
$ of the driving ac force is a multiple of the time $L/v$ during which the
soliton passes one spatial period: $2\pi /\omega =mL/v$,$\;$where $
m=1,2,3,...\;$is the order of the resonance. More general subharmonic
resonances, with
\begin{equation}
v=\left( m/M\right) \left( L\omega /2\pi \right) ,  \label{resonance}
\end{equation}
$m$ and $M$ being mutually prime integers, are possible too.

In the lattices, a similar effect was predicted in models of two different
kinds: Toda-type chains, in which the interaction between neighboring
particles is anharmonic (with no substrate potential), and lattices of the
Frenkel-Kontorova (FK) type, combining the harmonic interaction between the
neighbors and an anharmonic on-site potential. Note that the FK lattice
finds its straightforward (and, as a matter of fact, most realistic)
physical realization in the form of a an array of coupled small Josephson
junctions (traditionally, the FK model was applied to crystal-lattice
dynamics \cite{review}, where, however, this model may be too far idealized
because of the complexity and non-one-dimensionality of the real crystals,
strong dissipation and thermal fluctuations, etc.). Recently, precise
experiments with the Josephson arrays have begun \cite{Pedersen,array}. The
necessary drive is provided by the ac bias current uniformly applied to all
the junctions. Then, the progressive motion of the soliton-like excitation
can be easily observed in the form of mean dc voltage across the junctions.

In \cite{TL1}, the ac-driven motion of a dislocation was predicted
analytically for the Toda-lattice models, and in \cite{TL2} it was observed
in direct numerical simulations of an ac-driven weakly damped Toda lattice
(in its so-called dual form). Recently, a similar effect was described
analytically and found numerically also in a {\em parametrically} ac-driven
weakly damped Toda lattice \cite{TL3}.

In the earlier work \cite{BM}, the existence of a moving kink (lattice
dislocation) was predicted for the ac-driven FK model with weak friction
(cf. Eq.~(\ref{JJ})),
\begin{equation}
\ddot{\xi}_n-a^{-2}\left( \xi _{n-1}+\xi _{n+1}-2\xi _n\right) +\sin \xi
_n=-\alpha \dot{\xi}_n+\gamma \sin (\omega t)\,,  \label{FK}
\end{equation}
where $\xi _n(t)$ is the coordinate of the $n$-th particle in the lattice,
and the inverse coupling constant $a$ plays the role of the lattice spacing
in the quasicontinuum limit.
As well as in the continuum model (1), in Eq.\ref{FK} time is rescaled
so that the coefficient in front of the on-site nonlinear term
$\sin\xi_n$ is $\equiv 1$.  In the lattice, it is natural to measure the
travelled distance by the number of the sites that the soliton has
passed.  The accordingly defined velocity $u$ of the ac-driven soliton
in the lattice was predicted in \cite{BM} to be, in the general case,
\begin{equation}
u=\left( m/M\right) \left( \omega /2\pi \right) ,  \label{discres}
\end{equation}
where, as well as in Eq.~(\ref{resonance}), $m$ and $M$ are the super- and
subharmonic resonance orders.
Due to the evident reflectional symmetry of the lattice, the resonant velocity
may have either sign. In a physical or numerical experiment, the sign of the
velocity is determined by the initial conditions. Note also that the resonant
velocity \ref{discres} is not related to the phase or group velocity of
the free excitations in the ideal (undamped and undriven) linearized FK
lattice.

Along with the resonant velocity, an important characteristic of the
ac-driven motion is the threshold value $\gamma _{{\rm thr}}\left( \alpha
,\omega \right) $ of the drive's amplitude, i.e., a minimum value of the
amplitude which, at fixed values of $\alpha $, $\omega $ and other
parameters, admits the ac-driven motion. In the modulated continual model
(1), the threshold was found analytically in the framework of the
perturbation theory, treating $\epsilon $, $\alpha $, and $\gamma $ as small
parameters, and taking the kink solution to the unperturbed sine-Gordon (SG)
equation as the zero-order approximation . This analytical result was
demonstrated to be in good agreement with direct simulations \cite{Bob}. The
perturbation theory was also applied to the different versions of the
ac-driven damped Toda lattice, which had produced a prediction for the
threshold amplitude \cite{TL1,TL3} that was in a reasonable agreement with
the numerical simulations reported in \cite{TL2,TL3}.

For the ac-driven damped FK lattice, the situation is more difficult, as
even the unperturbed FK lattice, i.e., the one without the friction and
drive terms, is far from being solvable in any sense. Numerical simulations,
starting from the work \cite{PK}, have demonstrated that the unperturbed FK
lattice does not support any fully stable moving kink. Instead, the
discreteness gives rise to emission of radiation by the moving kink (which
may produce strong resonant effects \cite{rad}), i.e.,
an effective radiation friction force. Moreover, in certain cases
the emission of radiation may produce strong resonant effects \cite{rad}. In
any case, the ac drive in the model (\ref{FK}) must compensate both the
direct dissipative losses and radiation.

Because no well-defined zero-order approximation for the moving soliton is
available, the only practical possibility to get analytical estimates is to
use the quasicontinuum approximation, which was done in \cite{BM} in order
to predict the threshold for the existence of the ac-driven soliton in the
ac-driven damped FK lattice. However, the quasicontinuum approximation had
produced a threshold amplitude that was exponentially large in the
discreteness parameter $a$ (see Eq. (\ref{FK})).
In reality, the system may easily develop an instability and slide into a
chaotic regime before reaching such a large value of the driving amplitude
(see, e.g., \cite{wrong}).
Therefore, without direct
numerical simulations, it is not clear if one can rely upon the predictions
of the quasicontinuum approximation.

The first simulations of the kink's dynamics in the model (\ref{FK}) were
reported in \cite{wrong}, where no case of the ac-driven kink's motion was
found, and it had been concluded that this regime is impossible. In contrast
with that, a recent work \cite{strange} reported other results of
simulations of a FK lattice that consisted of 20 particles with $a=1$, the
ac-drive's period $T\equiv 2\pi /\omega $ took the values $50$ or
$25$, while the dissipative constant was $\alpha =0.1$. Note that the
corresponding friction-braking time $\sim 1/\alpha $ is essentially
smaller than the period $T$, hence this regime was, as a matter of fact,
{\em overdamped}. The most essential effect
observed in \cite{strange} was depinning of a kink trapped in the lattice
under the action of a rather strong
ac drive. In most cases, the overdamped motion of
the depinned kink was a diffusive drift; however, at some values of the
driving force, a nonzero mean velocity predicted by Eq.~(\ref{resonance})
with $m=M=1$ was observed. Thus, the results reported in \cite{strange}
confirm, in a limited case, the qualitative prediction made in \cite{BM}).
However, the basic characteristic of the ac-driven motion, viz., its
threshold, was not considered in \cite{strange},
and, in fact, the most interesting {\em weakly} damped case was
not at all dealt with in that work.

Our objective is to present results of systematic numerical simulations that
confirm the existence of the stably moving dislocation-like collective
excitations in the weakly damped ac-driven FK lattice with periodic boundary
conditions (b.c.), that include a phase jump of $2\pi $, necessary to
support the topological soliton. These collective excitations are found in
two very different forms: broad ones, that extend, roughly, to half the full
length of the periodic lattice (so that the term ``soliton'' is not
appropriate for them) and propagate keeping a virtually constant shape, and
narrow (really localized) quasi-solitons that are found in an {\em excited}
form, demonstrating persistent internal vibrations: the $2\pi $-kink
periodically splits into two $\pi $-subkinks that then recombine back into
it.

Because of the lack of a good initial guess, it proved to be virtually
impossible to generate these excitations directly. Therefore, it was
necessary to devise a special indirect procedure that made it possible to
generate the moving collective excitations, which is described in section 2.
In sections 3 and 4, we summarize the numerical results obtained,
respectively, for the broad and narrow moving dislocations. The threshold
characteristics for both types of the dislocations are presented in section
5, and the paper is concluded by section 6.

\section{THE MODE OF THE SIMULATIONS}

All the attempts to {\em directly} find the dynamical regime sought for in
the weakly damped model (\ref{FK}) have failed. A clue to successful quest
was to start from the case studied in the numerical part of the work \cite
{Bob}. Indeed, the simulations of the modulated continual model (\ref{JJ})
performed in that work amounted to a numerical solution of a {\em discrete}
system with the periodic modulation,
\begin{equation}
\ddot{\xi}_n-a^{-2}\left( \xi _{n+1}+\xi _{n-1}-2\xi _n\right) +\left[
1+\epsilon \sin \left( 2\pi n/N\right) \right] \sin \xi _n=-\alpha
\dot{\xi} _n+\gamma \sin (\omega t)\,  \label{discrete} \end{equation}
(cf. Eqs.~(\ref{FK})), subject to the periodic b.c. with the $2\pi $
phase shift, $\xi _N\equiv \xi _0+2\pi $. Note that the modulation
period introduced in Eq. (\ref{discrete}) exactly coincides with the
full size of the lattice.

It was easy to catch the moving $2\pi $-kink (dislocation) in simulations of
the modulated discrete system (\ref{discrete}). The initial displacements $
\xi _n(t=0)$ and the velocities $\dot{\xi}_n(t=0)$ were taken as per the
exact kink solution of the unmodulated continuum SG equation, moving at the
velocity
\begin{equation}
u_{{\rm in}}=N\omega /2\pi .  \label{initial}
\end{equation}
This velocity is suggested by the condition of the fundamental resonance
between the kink's motion and ac drive, which was our starting point.
Actually, the possibility to launch the moving dislocation in the modulated
system was not sensitive to details of the initial configuration, i.e., the
driven dislocation is a sufficiently strong attractor in the weakly damped
modulated system. Then, the simulations were continued, gradually decreasing
the modulation amplitude $\epsilon $ until it vanished. In some cases, the
ac-driven regime persisted up to $\epsilon =0$. The ac-driven regime of the
motion of the dislocation in the unmodulated lattice was regarded as stable
if it persisted for $\geq $ 10,000 circulations around the periodic lattice
(the corresponding time is larger than that sufficient for a freely moving
dislocation to be stopped by the friction force, typically, by a factor $
_{\sim }^{>}$ 500).

The study of the transition between the ac-driven regimes in the modulated
FK lattice (\ref{FK}) and the uniform one (\ref{discrete}) is of interest by
itself. Deferring a detailed consideration of this issue to another work, we
here demonstrate its most salient feature: keeping all the parameters but
the modulation amplitude $\epsilon $ constant, and gradually decreasing $
\epsilon $, we observed an abrupt decrease of the velocity $u$ by a jump at
a well-defined finite value of $\epsilon $ (Fig. 1). This feature proved to
be quite generic for the transition from the ac-driven motion in the
modulated lattice to the motion in the uniform one.

\section{THE BROAD DISLOCATIONS}

The first noteworthy numerical result for the uniform lattice ($\epsilon =0$
) is that a minimum value of the lattice size $N$, at which it was possible
to extend the ac-driven regime up to $\epsilon =0$, was $N_{\min }=12$,
while no upper limit for $N$ has been found (at least, up to $N=100$).
Another noteworthy peculiarity is that the ac-driven motion of the
dislocation could not be supported in the uniform FK lattice, unless the
driving frequency $\omega $ belonged to a narrow interval of a width $\Delta
\omega \sim 0.03$ around the value $\omega =1$. Therefore, all the plots
displayed below pertain to $\omega =1$.

A possible explanation to the latter feature can be provided by a {\em
cascading mechanism }described below. First, we note that $\omega =1$ is
exactly the {\it gap frequency} in the spectrum of the linear excitations, $
\xi _n\sim \exp \left( ikn-i\chi t\right) $, in the uniform FK model (\ref
{FK}), which is
\begin{equation}
\chi ^2(k)=1+4a^{-2}\sin ^2(k/2).  \label{spectrum}
\end{equation}
The spatially uniform ac drive with the frequency $\omega $, applied to the
lattice, excites uniform oscillations in it, which is the first step of the
cascading. At the next step, the modulational instability induced by the
nonlinearity stimulates decay of the uniform oscillations into a pair of
traveling waves with the same frequency $\omega $ and the wavenumbers $\pm k$
related to $\omega $ by the dispersion relation (\ref{spectrum}). Finally,
at the third step, the traveling wave can support ({\em drag}) a moving
dislocation. Note that dragging a dislocation (kink) by a given periodic
traveling wave was earlier considered in terms of other physical models
(see, e.g., \cite{drag}). If $\omega $ is close to the gap frequency of the
lattice, the generated traveling waves have a large wavelength $2\pi
/k\approx 2\pi /a\sqrt{\omega -1}$ (provided that $\omega >1$), and a
relatively large amplitude $\sim \left[ (\omega ^2-1)^2+\alpha ^2\right]
^{-1/2}$. This may be an explanation for the fact that the moving
dislocations supported by the ac drive have a large size (see below), and
that they can be supported only when $\omega $ is sufficiently close to $1$.

In the established regime obtained at $\epsilon =0$ by means of the
procedure described above, its {\it commensurability index} $r$ was found
from the numerical data, using Eq.~(\ref{discres}) as the definition:
\begin{equation}
r\equiv 2\pi u/\omega ,  \label{r}
\end{equation}
where $u$ is realized as the mean velocity of the ac-driven soliton. Thus
found value of $r$ was always quite high, but smaller than its large initial
integer value $r_{{\rm in}}\equiv N$ in the modulated lattice (see Eq.~ (\ref
{initial})), which is explained by the abrupt drop of $u$ seen in Fig. 1.
Moreover, the value of $r$ found at $\epsilon =0$ turns out to be {\em non}
integer. In Fig. 2, we display the plot of $r$ vs. $a$, obtained at fixed
values $N=25$, $\omega =1$, $\gamma \simeq 0.5$, and $\alpha =0.005$. It is
necessary to stress that, at $N=25$, it was not possible to find the stable
ac-driven regime at $\epsilon =0$ outside the limited range of the values of
$a$ shown in Fig. 2.

At other values of the parameters $N$, $\gamma $, and $\alpha $, the
dependencies $r(a)$ were found to be quite similar to those shown in Fig. 2.
In particular, for each value of the lattice size $N$, there is its own
interval of the values of $a$, in which the ac-driven motion can be
observed. Defining {{\bf $\bar{a}$ }}as the central point of this interval,
we were able to make the following observation: the product $\bar{a}N$
changes only from $5.25$ to $5.28$ for $N$ varying between $25$ and $100$. A
straightforward implication of this observation is that, in the
quasicontinuum limit, the length $L\equiv aN$ of the closed system
admitting
the ac-driven motion of the dislocation, is nearly constant (recall,
however, that the ac-driven motion of the dislocation(kink) is
impossible in
the uniform continual system).

Actually, the commensurability index is {\em irrational}, which is seen from
a typical phase portrait of the ac-driven regime displayed in Fig. 3a. To
generate the phase portrait, an arbitrary lattice site was selected, and on
its phase plane ($\xi _n$,$\dot{\xi}_n$), a discrete trajectory was plotted,
consisting of the points picked up at $t=\left( 2\pi /\omega \right) j$,
where $j$ is discrete time taking integer values. The quasicontinuous
evolution of $\xi _n$ obvious in Fig. 3a implies that the commensurability
index is irrational indeed. Note, however, that the phase portraits show no
trace of dynamical chaos. Simultaneously, they strongly suggest that, on the
phase plane ($\xi _n$, $\dot{\xi}_n$), there must exist two {\em fixed points
}, one being a saddle corresponding to the self-intersection of the
trajectory, the other one being a central point inside the loop formed by
the self-intersecting trajectory. The fixed points correspond to $r=N$,
i.e., the solution in which the soliton completes exactly one round trip in
the periodic lattice during one period of the external drive. However, we
have not been able to capture the corresponding solutions in the
simulations. It is obvious that the saddle is an unstable fixed point, hence
it cannot be found in the dynamical simulations. As concerns the central
fixed point, it would be stable in a conservative system, but it may easily
become unstable in our dissipative driven model, which is a plausible
explanation for the impossibility of capturing this fixed point.

Although the actually found solution of the present type is nonstationary,
its nonstationarity is mild: the dislocation does not manifest conspicuous
internal vibrations, and, at any moment of time, its shape is close to that
in Fig. 3b, which is an instantaneous snapshot showing $\dot{\xi}_n$ vs. the
lattice site number $n$, taken at an arbitrarily chosen moment of time. Note
that the presentation of the shape in this form  is relevant for the arrays
of coupled small Josephson junctions \cite{array}: $\dot{\xi}_n$ is
the instantaneous voltage at the $n$-th junction.

As one sees from Fig. 3b, the moving excitation is broad, extending,
roughly, to half the full length of the lattice. Note, however, that this
excitation is, definitely, a dislocation, as it bears the topological charge
$2\pi $. In view of the fact that this moving dislocation is observed at the
values of the discreteness parameter $a\simeq 0.2$ (see Fig. 2), and the
size of the dislocation is $_{\sim }^{>}\,10$ (Fig. 3b), one may conjecture
that the dislocation should have its counterpart in the corresponding
continual model, i.e., the ac-driven damped SG equation. However, we recall
once again that the progressive motion of a collective excitation is
impossible in the ac-driven continual systems. Alternatively, one can assume
that the moving dislocation is a $2\pi $ kink massively ``dressed'' by the
long-wave lattice phonons, which would comply with the above-mentioned
cascading mechanism, that may feasibly account for the energy transfer from
the ac drive to the dislocation. Much more work is necessary to
elaborate this possibility.

\section{THE NARROW DISLOCATIONS AND THEIR VIBRATIONS}

The broad dislocations considered in the previous section can hardly be
called ``solitons''. However, a dislocation of another type, which is much
better localized, can also be found. To this end, we start from the
modulated FK model \ref{discrete}, with the periodic b.c. corresponding
to the lattice size $2N$ (instead of the size $N$ dealt with above), so
that the initial dislocation, being essentially the same as in the
previous section, is twice as narrow in comparison with the full lattice
size. Then, the procedure of decreasing the modulation amplitude
$\epsilon $ to zero produces a nondecaying excitation essentially
different from that considered in the previous section. As well as the
broad dislocations, the new ones were observed only in a narrow interval
of the frequencies $\Delta \omega \simeq 0.03$ around $\omega =1$.

A typical phase portrait of the dislocation of this type is shown in Fig. 4,
which is obviously different from Fig. 3a. The moving dislocation
demonstrates strong persistent internal vibrations. Approximately
periodically, it oscillates between two configurations, one of which
corresponds to a clearly localized $2\pi $-kink, and the other one to a set
of two separated $\pi $-subkinks (Fig. 5). We stress that the internal
vibrations of the narrow dislocation (as well as its progressive motion)
continue indefinitely long, being, obviously, permanently supported by the
ac drive. A numerically calculated temporal Fourier transform of the
solutions (not shown here) demonstrates that the frequency of the internal
vibrations of the narrow dislocation moving at the mean velocity $u$ is
close to {\bf $au$. }It is necessary to mention that the kink in the
unperturbed FK lattice is known to have an {\em internal vibrational mode}
\cite{review,braun97}, which may
be quite a natural explanation for the
persistent internal oscillations of the narrow dislocation.
The internal mode can be easily excited by the periodic perturbation exerted by
the lattice on the dislocation moving past it. Of course, the progressive
motion of the dislocation may be essentially affected by the interaction of the
corresponding degree of freedom with the internal vibrations. In this work, we
do not consider the latter issue in detail.

Thus, this type of the moving collective excitation can be easily generated
only in a vibrating state. On the other hand, the fixed points corresponding
to the centers of the three loops seen in Fig. 4 (two upper and one lower)
should represent this dislocation in its ground state. Though it is
virtually impossible to directly generate the dislocation in this state, its
shape can be effectively restored. To this end, we picked up a number of
points belonging to one of the three loops and distributed uniformly along
it. A presumably stationary shape of the narrow dislocation in its ground
state was produced by juxtaposing and averaging the configurations taken at
the selected points belonging to the loop. The result is shown in Fig. 6. As
one sees in this figure, the ground state is indeed a narrow soliton-like
dislocation, whose width is about $1/4$ of the full lattice size. The
``tail'' attached to this narrow dislocation may, in principle, be either an
artifact generated by the approximation used, or a genuine feature produced
by the long-wave lattice phonons dressing the dislocation. We stress that
the juxtaposing and averaging procedure has produced, with a reasonable
accuracy, the {\em same} result when applied to all the three loops in Fig.
4. Because the above-mentioned ``tail'' was reproduced virtually in the same
form in all the three cases, we conjecture that it is a genuine feature of
the narrow dislocation, rather than an artifact.
Accurate examination of the internal pulsations of the narrow dislocation
shows that the pulsations are quite irregular (probably, chaotic). They
have, in approximate sense, sort of a basic period (not quite constant),
which is close to six periods of the ac drive. Thus, there is no obvious
resonance between the internal oscillations of the narrow dislocation and the
external drive. Instead, this seems like a typical portrait of chaotic
oscillations in a weakly damped nonlinear dynamical system driven by a
periodic external force (as is well known, the dynamical regime in such
a system should be, generally, chaotic, even if the system has a single
degree of freedom).

A natural question is whether more types of the dislocations can be produced
by a generalization of the procedure that gave rise to the narrow
dislocation, starting with the modulated lattice of the full size $3N$,
$4N$, etc. The answer to this question is {\em negative}: at least,
in the case
of the full size $3N$ and $4N$, no new dislocation could be generated.
So we just have the cases $N$ and $2N$, this is too little to infer on the
infinite limit. The fundamental question if the dislocation we have
investigated is a soliton or not cannot be answered satisfactorily with
the simulation here presented. However, in the two cases
investigated, the head of the excitation is still interacting with the
tale, so we should for the moment conclude that it is not a soliton.

\section{THE THRESHOLD CHARACTERISTICS}

As it was mentioned above, a basic characteristic of the ac-driven regime is
its threshold, i.e., a minimum value $\gamma _{{\rm thr}}$ of the drive's
amplitude $\gamma $ that allows one to support the motion of the dislocation
at a nonzero mean velocity. Because the driving force must (first of all)
compensate the friction, we present, in Fig. 7, $\gamma _{{\rm thr}}$ vs.
the friction constant $\alpha $ at the driving frequency $\omega =1$ and
different values of the lattice size $N$, selecting the values of the
inverse coupling $a$ so that to have a fixed value of the soliton's velocity
$v=au$, defined as per the quasicontinuum approximation, in each plot. The
data are presented for both types of the dislocations, broad and narrow. A
remarkable feature is that the plots $\gamma _{{\rm thr}}\left( \alpha
\right) $ for the dislocations of the former type are nearly universal, very
weakly depending upon the lattice size $N$ (while the size of the broad
dislocation strongly depends on $N$, being $\simeq N/2$ according to the
above results). It is also noteworthy that, for the narrow dislocations, the
threshold is essentially lower, which may be related to the fact that, on
average, this dislocation involves into the collective motion fewer
particles in the lattice, hence the dissipative losses are smaller.

A straightforward perturbation theory predicts the dependence $\gamma _{{\rm
thr}}\left( \alpha \right) $ to be linear \cite{Bob,TL1,BM}. The
dependencies displayed in Fig. 7 are, in fact, not far from being linear,
but an extrapolation (with a first order polynomial best fit) suggests
that $\gamma _{{\rm thr}}$ remains different from zero at $\alpha = 0$.
A natural explanation for this is that the driving force must compensate
not only the direct dissipative losses, but also additional losses
induced by the emission of radiation \cite{rad}.  The internal mode can
be easily excited by the periodic perturbation exerted by the lattice on
the dislocation moving past it. Of course, the progressive motion of the
dislocation may be essentially affected by the interaction of the
corresponding degree of freedom with the internal vibrations. In this
work, we do not consider the latter issue in detail.

Thus, at
very small $\alpha $, the radiation losses may dominate, demanding a finite
drive's amplitude in the limit $\alpha \rightarrow 0$. Indeed, at extremely
low values of the damping ($\alpha$ below $\approx 0.001$, the lowest
damping shown in Fig. 7), the radiation processes seem to
dominate in the simulations.
Actually, the ac-driven motion of the dislocation is very difficult to
capture in the case of extremely weak dissipation, i.e., the dissipation
is a stabilizing factor for the ac-driven dislocation (as long as the
system does not become overdamped).

\section{CONCLUSION}

In this work we have demonstrated, by means of direct simulations, that two
species of moving collective nonlinear excitations of the dislocation type
(distinguished by the topological charge $2\pi $) may exist in the ac-driven
weakly damped FK lattice with the periodic boundary conditions. Direct
generation of the moving solitons turned out to be impossible; however, they
can be generated initially in the lattice subject to a spatial modulation of
the on-site potential strength. Then, gradually decreasing the modulation
depth, we were able to find stable moving dislocations in the uniform FK
lattice, provided that the lattice size is $N\geq 12$, and the driving
frequency is close to the gap frequency of the linearized model.

All the collective excitations that were found have a large noninteger
index of commensurability with the lattice. The results
suggest that the exact value of the commensurability index is likely to be
irrational. The moving
dislocations of two different types have been found. The first type
represents broad dislocations with a nearly stationary shape, extending to
approximately half of the whole length of the lattice. The other type is
represented by the collective excitations that, when generated directly,
demonstrate strong persistent internal vibrations between the configurations
corresponding, respectively, to a relatively narrow $2\pi $-kink and two
well separated $\pi $-kinks. A shape of the dislocation of the latter type
in the ground (nonvibrating) state was recovered indirectly, by means of
juxtaposing and averaging many configurations picked up from a loop
surrounding the corresponding fixed point.. This ground state proves to be a
soliton-like $2\pi $-kink whose width is about $1/4$ of the lattice size.

The threshold amplitude of the driving force was found as a function of the
friction constant $\alpha $. An extrapolation of this dependence suggests
that a nonzero drive's amplitude remains necessary at $\alpha \rightarrow 0$
, which may be explained by radiation losses. However, the ac-driven
dislocation becomes virtually unstable at vanishingly small values of $
\alpha $.

The effects studied in this work theoretically can be easily observed in an
array of linearly coupled small Josephson junctions, in the form of dc
voltage generated by the ac bias current uniformly applied to the array.
Note that , once the shape of the moving dislocation in the uniform lattice
has been found, the corresponding initial configuration can be easily
generated in the experiment, hence there is no real need in the auxiliary
spatial modulation of the lattice, that was a crucial trick in the
simulations.

Although qualitative explanations of some features of the observed effects were
put forward in this work, full understanding of the ac-driven motion of
dislocations in weakly damped nonlinear lattices is still lacking.

\section*{ACKNOWLEDGMENTS}

We appreciate useful discussions with D. Cai and O.M. Braun. This work was
strongly supported, at its initial stage, by R.D. Parmentier, who tragically
died January 2, 1997.

%\newpage

\begin{center}
%\newpage
{\bf {Figure Captions} }
\end{center}

Fig. 1. A typical example of the abrupt decrease of the ac-driven soliton's
velocity in the modulated FK lattice with the decrease of the modulation
amplitude $\epsilon $. In this figure, $\gamma =0.5$, $a=0.0564$, $\alpha
=0.005$, and $N=100$. In this and all the other figures, $\omega =1$.

Fig. 2. The effective commensurability index defined as per Eq. (\ref{r})
vs. the inverse harmonic coupling constant $a$, as obtained from the
numerical data. Because $\omega =1$, this plot, as a matter of fact, also
shows the velocity of the moving dislocation. The other parameters are
$\gamma \simeq 0.5$, $\alpha =0.005$, and $N=25$.

Fig. 3. Typical examples of
the phase portrait (a) and the shape (b) of the broad ac-driven dislocation
at $\gamma =0.5$, $a=0.226$, $\alpha =0.005$, and $N=25$. The mean velocity
of this dislocation is $u=3.10$.

Fig. 4. A typical example of the phase portrait of the narrow ac-driven
dislocation at $\gamma =0.5$, $a=0.226$, $\alpha =0.005$, and $2N=50$.
The mean velocity of this dislocation is $u=2.80$.

Fig. 5. Three configurations between which the narrow dislocation vibrates:
a $2\pi $-kink (short-dashed line), the two separated $\pi $-subkinks (solid
line), and an intermediate profile (long-dashed line). The values of the
parameters are the same as in Fig. 4, except for the actual size of the
lattice which, according to the procedure adopted for the generation of the
narrow dislocations, is $2N=50$.

Fig. 6. The indirectly retrieved configuration of the narrow dislocation in
its ground state: (a) $\xi _n$ vs. the discrete coordinate $n$; (b) $\dot{\xi
}_n$ vs. $n$. The values of the parameters are the same as in Fig. 5.

Fig. 7. The dependencies $\gamma _{{\rm thr}}\left( \alpha \right) $, the
value of $a$ being selected so that the soliton's velocity $v\equiv au$,
defined as per the quasicontinuum approximation, keeps a constant value, $
v=0.7$. The stars, crosses, vertical scores, and rhombuses pertain,
respectively, to the broad dislocations in the lattices with the lengths,
respectively, $N=12$, $N=25$, $N=50$, and $N=100$. The David's shields
represent the narrow dislocations in the lattice with $2N=50$.


\begin{thebibliography}{99}
\bibitem{general}  {\it Microscopic Aspects of Nonlinearity in Condensed
Matter}, ed. by A.R. Bishop, V.L. Pokrovsky, and A. Tognetti (Plenum: New
York, 1991).

\bibitem{review}  O.N. Braun and Yu.S. Kivshar, Phys. Rep. {\bf 306}, 1
(1998).

\bibitem{MS}  D.W. McLaughlin and A.C. Scott, Phys. Rev. A {\bf 18}, 1652
(1978).

\bibitem{Bob}  G. Filatrella, B.A. Malomed, and R.D. Parmentier, Phys. Lett.
A {\bf 198}, 43 (1995).

\bibitem{Ustinov}  I.L. Serpuchenko and A.V. Ustinov, Solid State Comm. {\bf
68}, 693 (1988); A.A. Golubov, A.V. Ustinov, and I.L. Serpuchenko, Phys.
Lett. A {\bf 130}, 107 (1988).

\bibitem{Pedersen}  N.F. Pedersen and A.V. Ustinov, Supercond. Sci. Technol.
{\bf 8}, 389 (1995).

\bibitem{array}  S. Watanabe, S.H. Strogatz, H.S.J. van der Zant, and T.P.
Olrando, Phys. Rev. Lett. {\bf 74}, 379 (1995).

\bibitem{TL1}  B.A. Malomed, a paper in \cite{general}, p. 159, and Phys.
Rev. A {\bf 45}, 4097 (1992).

\bibitem{TL2}  T. Kuusela, J. Hietarinta, and B.A. Malomed, J. Phys. A {\bf
26}, L21 (1993).

\bibitem{TL3}  K.D. Rasmussen, B. A. Malomed, A.R. Bishop, and N. Gr\o
nbech-Jensen, Phys. Rev. E {\bf 58}, 6695 (1998).

\bibitem{BM}  L.L. Bonilla and B.A. Malomed, Phys. Rev. B{\bf \ 43}, 11539
(1991).

\bibitem{PK}  M. Peyrard and M.D. Kruskal, Physica D {\bf 14}, 88 (1984).

\bibitem{rad}  A.V. Ustinov, M. Cirillo, and B.A. Malomed, Phys. Rev. B {\bf
47}, 8357 (1993).

\bibitem{wrong}  D. Cai, A. S\'{a}nchez, A.R. Bishop, F. Falo, and L.M.
Flor\'{i}a, Phys. Rev. B {\bf 50}, 9652 (1994).

\bibitem{strange}  P.J. Mart\'{i}nez, F. Falo, J.J. Mazo, L.M. Flor\'{i}a,
and A. S\'{a}nchez, Phys. Rev. B {\bf 56}, 87 (1997).

\bibitem{drag}  B.A. Malomed, J. Phys. Soc. Jpn. {\bf 62}, 997 (1993).


\bibitem{braun97} O.M. Braun, Y.S. Kivshar, and M.Peyrard, Phys. Rev.
{\bf E56}, 6050 (1997).

\end{thebibliography}
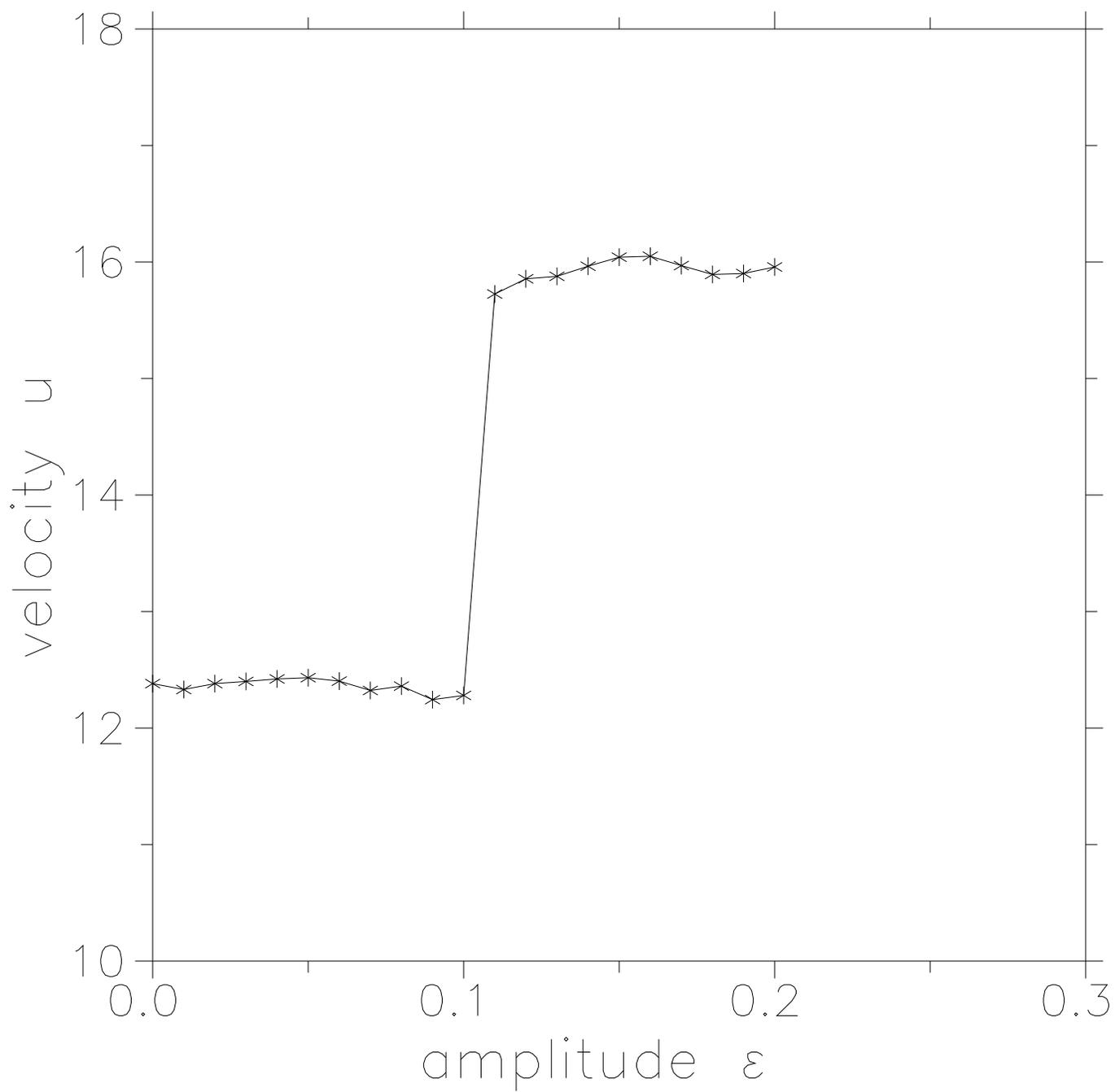
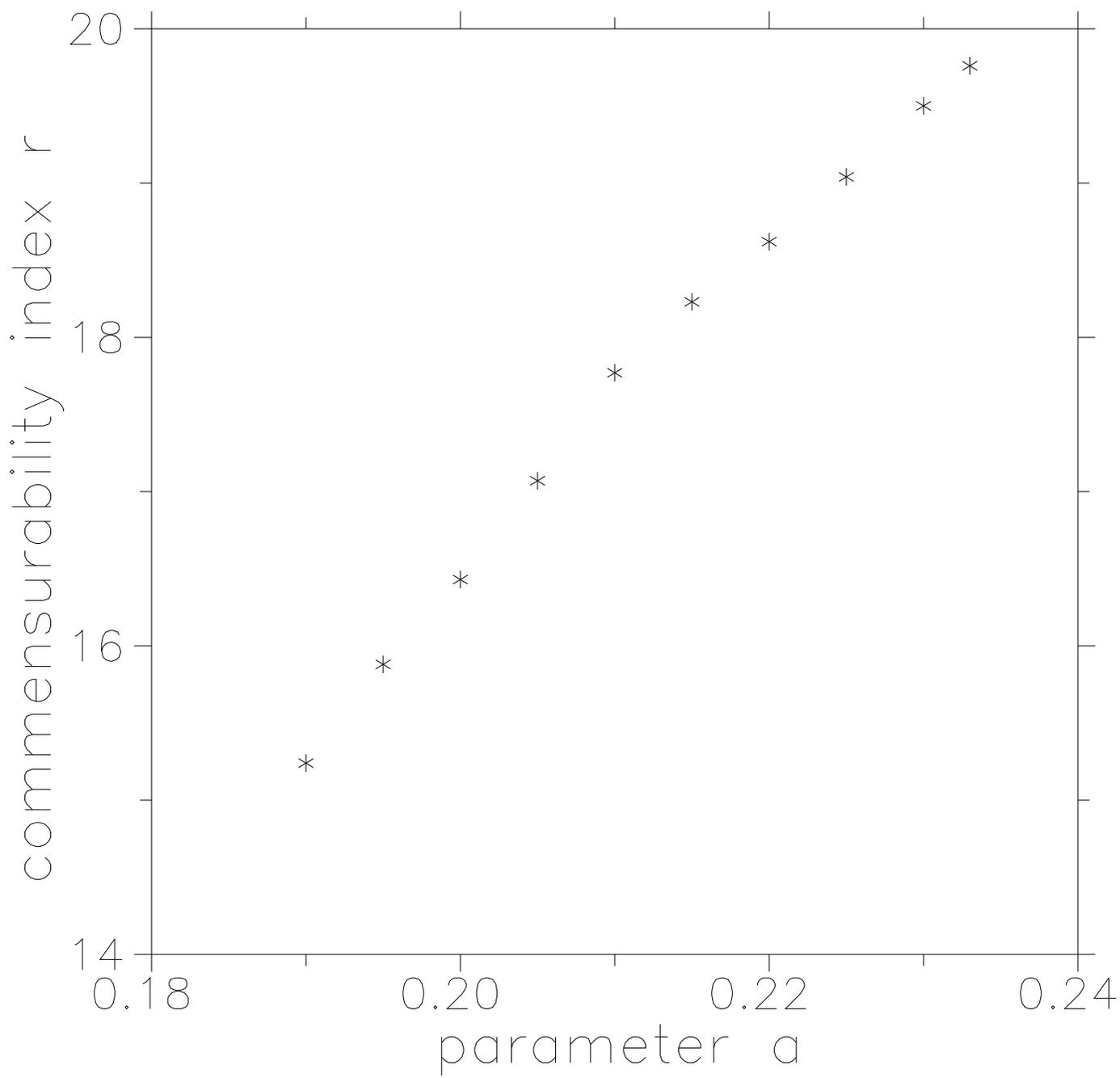
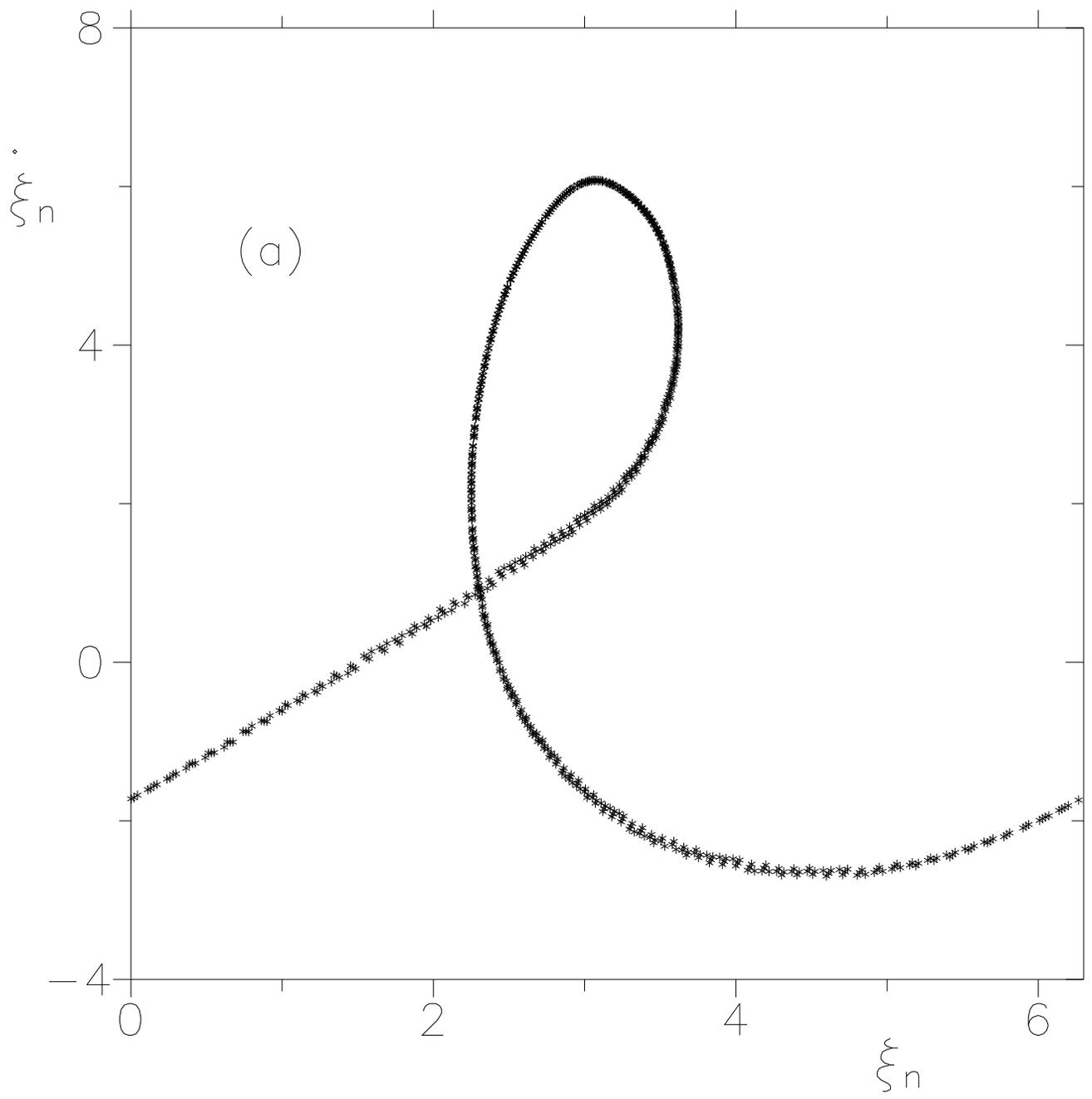
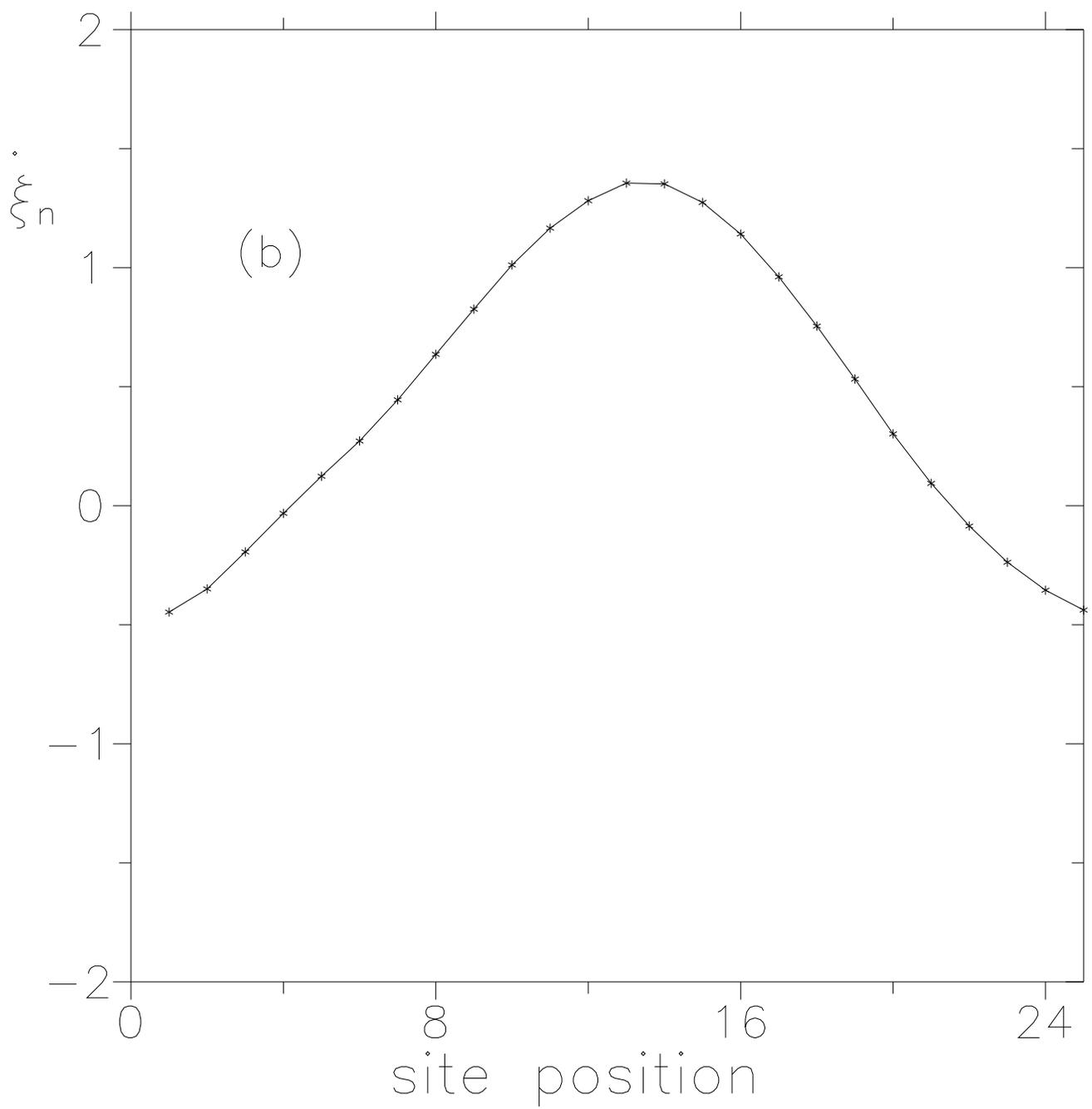
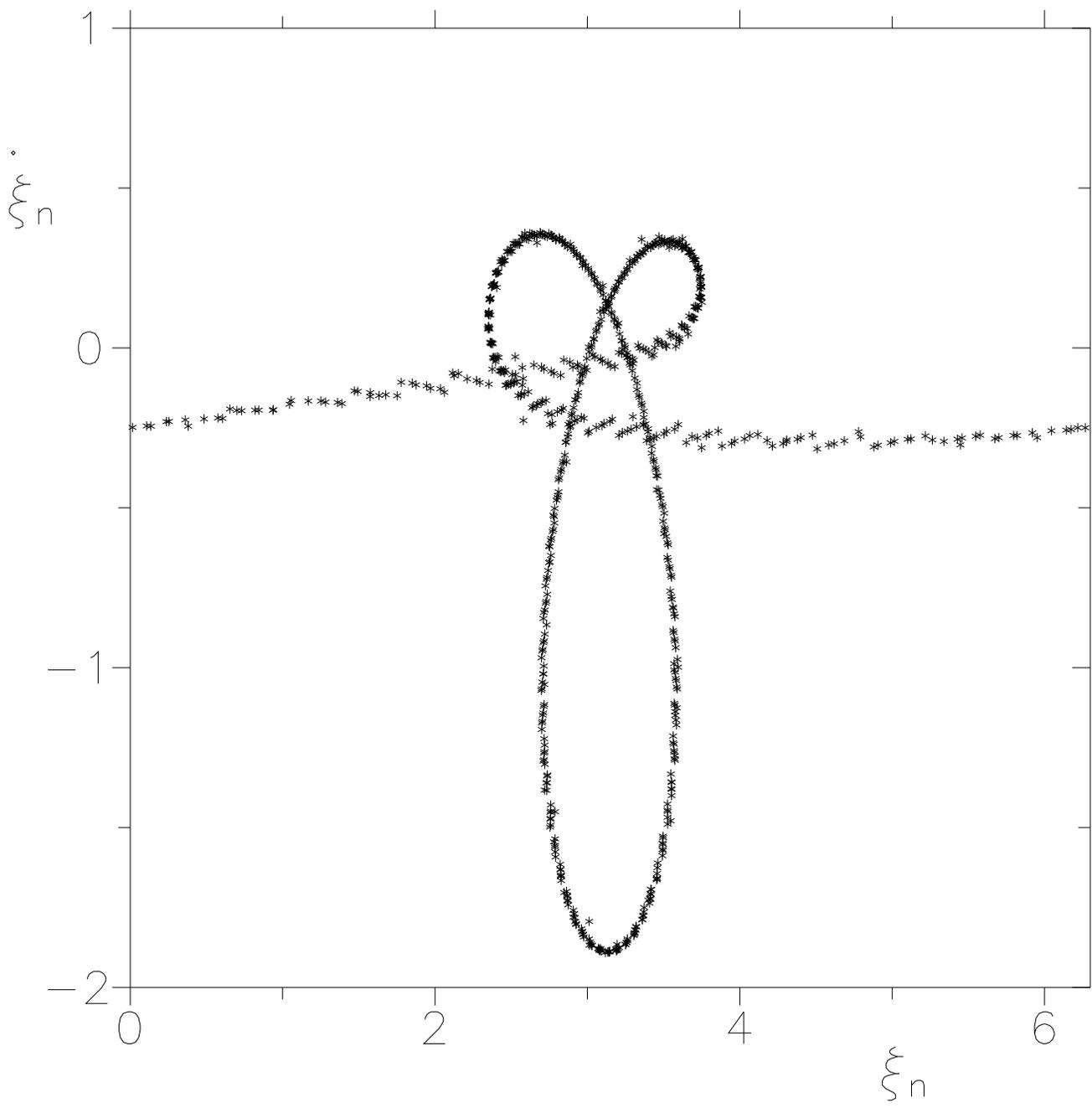
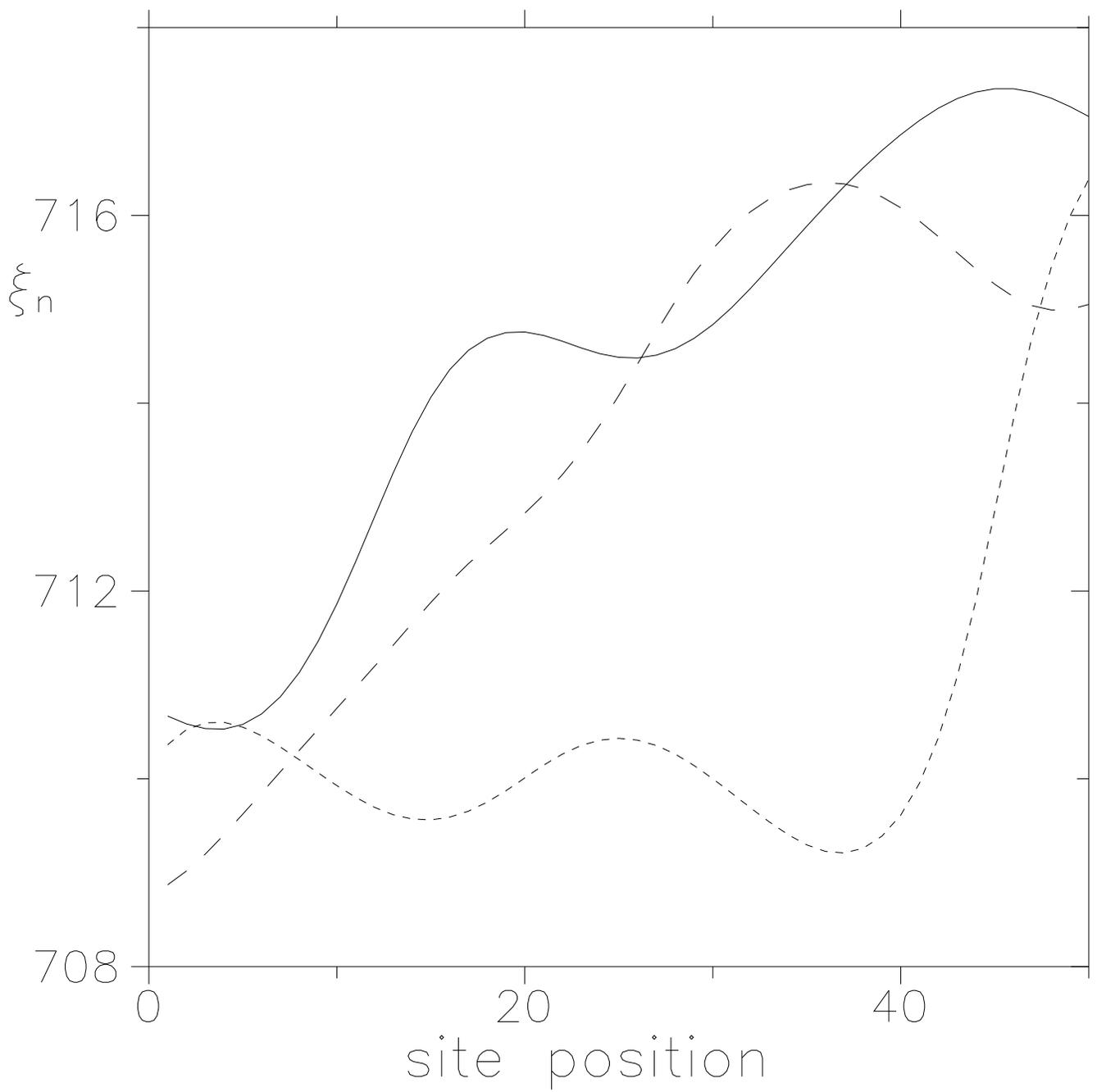
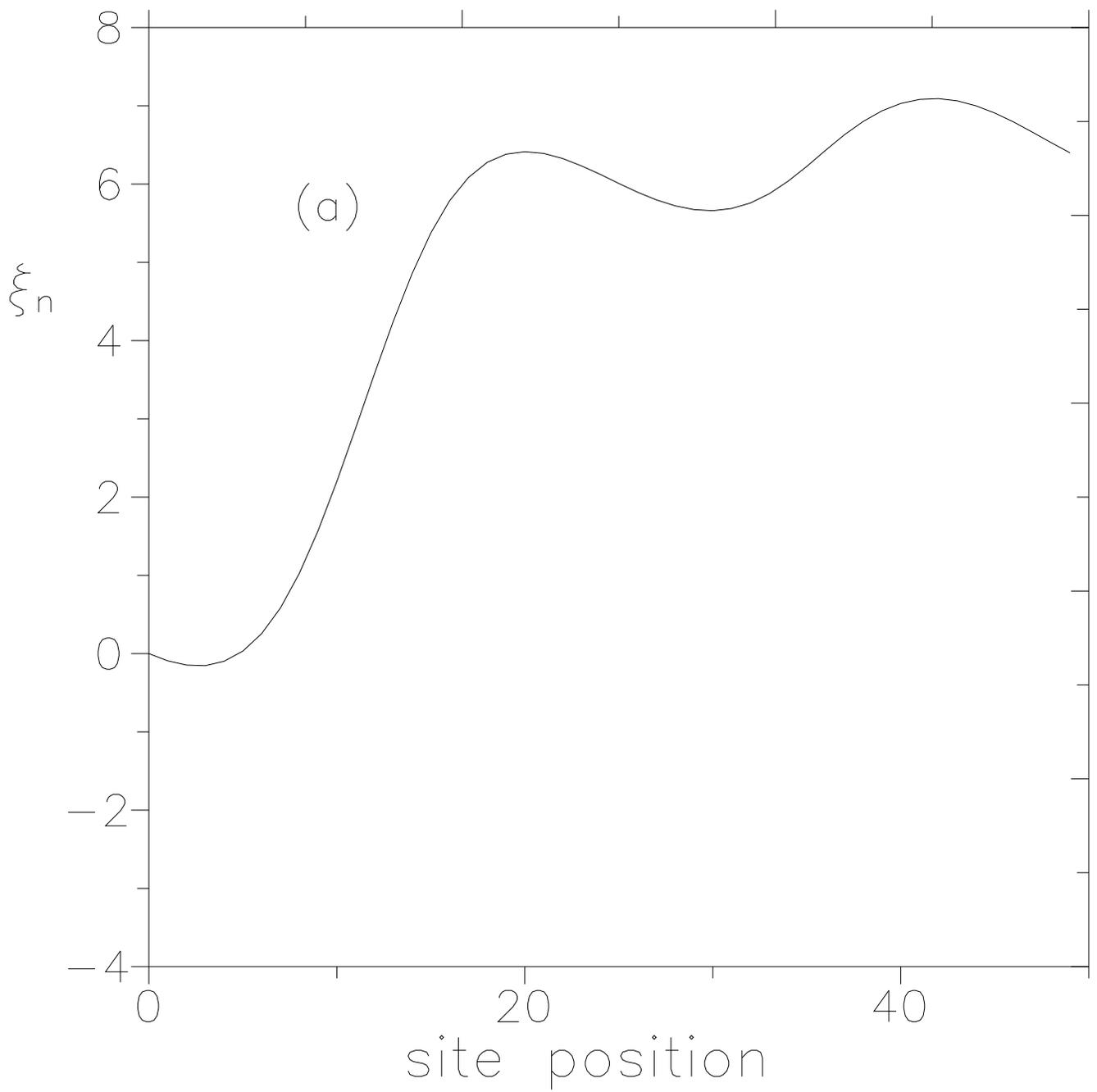
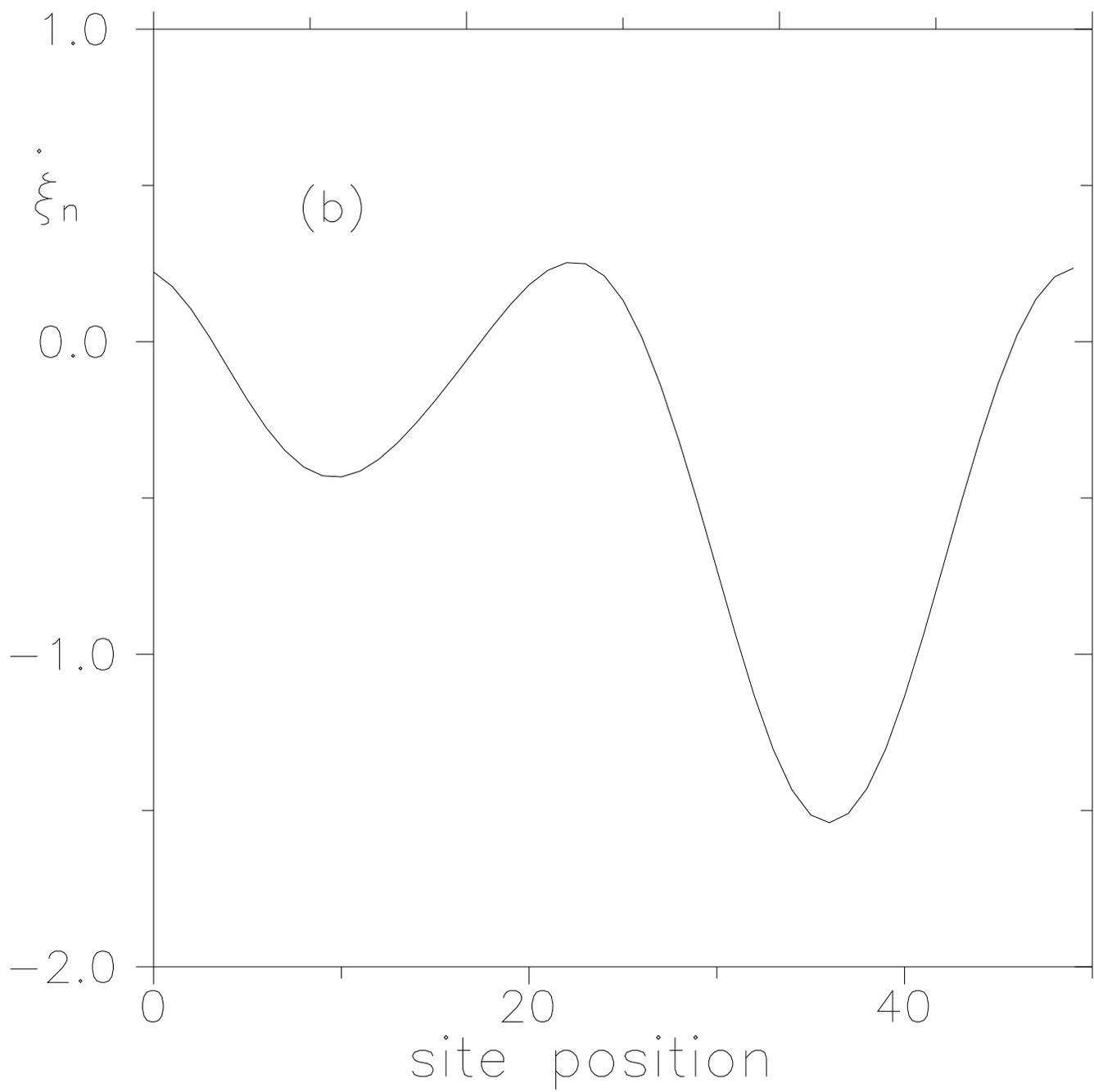
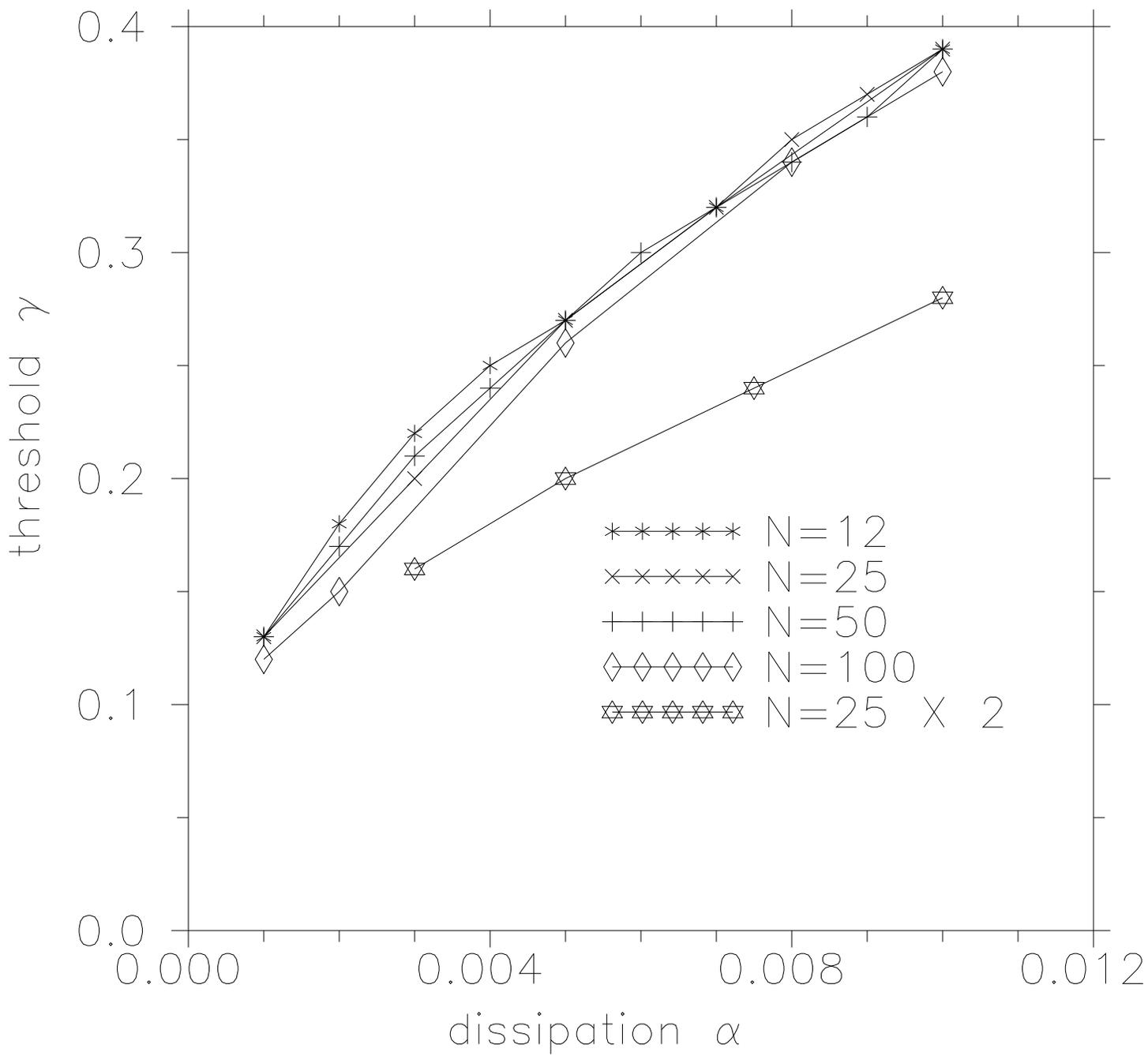
\end{document}